\documentstyle[aps,preprint]{revtex}
\newcommand{\beq}{\begin{equation}}
\newcommand{\eeq}{\end{equation}}
\newcommand{\beqa}{\begin{eqnarray}}
\newcommand{\eeqa}{\end{eqnarray}}
\newcommand{\n}{\nonumber\\}

\newcommand{\cV}{{\cal V}}
\newcommand{\hcV}{\widehat{\cal V}}
\newcommand{\hG}{\widehat{G}}
\newcommand{\hE}{\widehat{E}}
\newcommand{\hB}{\widehat{B}}
\newcommand{\hA}{\widehat{A}}
\newcommand{\D}{\Delta}
\newcommand{\E}{{\cal E}}
\begin{document}
\preprint{DESY 95-126}
\vfill
\title{
Colliding Wave Solutions, Duality, and Diagonal Embedding
of General Relativity in Two-Dimensional Heterotic String Theory}
\author{Shun'ya MIZOGUCHI
\footnote{Electronic address: mizoguch@x4u2.desy.de
}
}\address{
II. Institut f\"{u}r Theoretische Physik,
            Universit\"{a}t Hamburg \\
            Luruper Chaussee 149, 22761 Hamburg, Germany.
}
\date{June, 1995}
\maketitle
\begin{abstract}
The non-linear sigma model of the dimensionally reduced Einstein
(-Maxwell) theory is diagonally embedded into that of the two-dimensional
heterotic string theory. Consequently, the embedded string backgrounds
satisfy the (electro-magnetic) Ernst equation. In the pure Einstein
theory, the Matzner-Misner SL(2,{\bf R}) transformation can be viewed as
a change of conformal structure of the compactified flat two-torus,
and in particular its integral subgroup SL(2,{\bf Z}) acts as the modular
transformation. The Ehlers SL(2,{\bf R}) and SL(2,{\bf Z}) similarly act
on another torus whose conformal structure is induced through the
Kramer-Neugebauer involution. Either of the Matzner-Misner and the Ehlers
SL(2,{\bf Z}) can be embedded to a special T-duality, and if the former
is chosen, then the Ehlers SL(2,{\bf Z}) is shown to act as the S-duality
on the four-dimensional sector. As an application we obtain some new
colliding string wave solutions by using this embedding as well as
the inverse scattering method.
\end{abstract}
%\pacs{}
\newpage

\section{Introduction}
Recently duality symmetries have attracted much attention
in string theory and its low-energy effective field theory.
The symmetries of interest consist of two different types
of discrete groups --- T- and S- duality groups. The former
reflects an automorphism of the target space (See \cite{Tduality}
for a review.), while
the latter is a conjectured non-perturbative weak-strong coupling
duality symmetry \cite{Sduality}
in the sense of the string-loop expansion. It has been argued that
this may also be an exact symmetry of a full string theory
when the theory has enough ($N=4$) supersymmetries
so that a certain kind of non-renormalization theorem may ensure to
extend the tree-level result to the full theory
(See \cite{Senrev,Schwarzrev}).

A characteristic feature of the duality symmetries is
that they are discrete subgroups of non-compact symmetries
of effective supergravity theories known for a long time.
For example, in the toroidal compactification
to four dimensions the low-energy effective action
of heterotic string theory
%(at the generic point in the moduli space)
\cite{Narain}
not only exhibits a manifest O(6,22), but also possesses
an SL(2,{\bf R}) symmetry \cite{CSF}.
The Dynkin diagram of this group gets an additional node to be O(8,24)
in the further reduction to three dimensions \cite{MarcusSchwarz,Sen3d},
and, in accord with the general rule \cite{Julia}
of the dimensional reduction of nonlinear sigma models,
gets one more in two dimensions to be the loop group
$\widehat{\rm O}$(8,24), in which the duality group is embedded
\cite{Sen2d}. The integrable supergravity theories arising in this way
have already been studied in detail in Refs.\cite{Nicolai,Nicolai2}.
The type II string version, which fits in the
``hidden'' E$_7$ symmetry \cite{E7} in the case of four dimensions,
was also investigated and called ``U-duality''
\cite{HullTownsend}.

The similarity between the appearance of hidden symmetries in
supergravity theories and the Geroch group \cite{Geroch} in general
relativity was first pointed out in Ref.\cite{Julia}, and was
elaborated in \cite{BM}. It is well-known
that the reduced Einstein gravity to two dimensions possesses
(in the infinitesimal form) the affine $\widehat{\mbox{sl}}$(2,{\bf R})
Kac-Moody symmetry with a non-trivial central term \cite{Julia}.
It is generated by two fundamental finite
sl(2,{\bf R}) subalgebras, which are called
the Matzner-Misner \cite{MM} and the Ehlers
\cite{Ehlers} sl(2,{\bf R}), respectively. The Matzner-Misner
SL(2,{\bf R})
--- a general-relativity analogue of the T-duality (if restricted to
its integral discrete subgroup)--- is a symmetry of the
SL(2,{\bf R})/U(1)
nonlinear sigma model obtained in the dimensional reduction of the
Einstein action directly from four to two dimensions. The Ehlers
SL(2,{\bf R}) --- an analogue of the S-duality --- is a symmetry of
another SL(2,{\bf R})/U(1) nonlinear sigma model which is constructed
by trading the Kaluza-Klein vector field in three dimensions with the
axion-like ``twist potential''. The purpose of this paper is
to make the relation between the two theories
more concrete by ``diagonally'' embedding
the two-dimensional nonlinear sigma model associated with the
dimensional reduction of the Einstein(-Maxwell) theory into that
of the two-dimensional effective heterotic string theory
compactified on an eight-torus.

It is obvious that such an embedding is possible.
Specifically, we will diagonally embed the SL(2,{\bf R}) as a subgroup
of ${\rm O}(8,8)\supset {\rm O}(2,2)\otimes I_4$ (where $I_n$ denotes
the $n$-dimensional identity matrix),
and further the SU(2,1), the symmetry group of the reduced
Einstein-Maxwell theory, as a subgroup of
${\rm O}(8,24)\supset {\rm O}(2,6)\otimes I_4$. We consider this
embedding as we are motivated by the following two practical relevances.
First, by this embedding we are endowed with a simple picture
of the T- and S- duality symmetries on the particular string backgrounds.
We will show that the integral Matzner-Misner SL(2,{\bf Z})
transformation becomes a special T-duality, while the integral
Ehlers SL(2,{\bf Z}) acts as the S-duality on the four-dimensional
sector, if they are embedded into the string theory.
On the other hand, it can be shown that the Matzner-Misner SL(2,{\bf R})
transformation is a change of conformal structure of the
``compactified'' flat two-torus, and in particular the integral
ones are the modular transformations. Correspondingly, the Ehlers
SL(2,{\bf R}) similarly changes the conformal structure of another torus
induced through the Kramer-Neugebauer involution.
Thus it provides us a simple picture of the T- and S- duality
groups acting on the embedded configurations. Second, more
importantly, since the embedding is diagonal,
the sigma model Lagrangian of the effective two-dimensional
heterotic string on the embedded special backgrounds takes completely
the same form as that of the reduced Einstein(-Maxwell) theory. This
means that the embedded string backgrounds satisfy the (electro-magnetic)
Ernst equation. Hence one may recast every known solution of the
(electro-magnetic) Ernst equation into that of the corresponding
string background. We would like to emphasize here the difference
between our method and the
solution-generating techniques applied to string theory in the
previous litrature, in the latter of which
one takes a known solution of the
Einstein theory as a {\em trivial} solution of string theory with
all other fields set to zero, and then apply the duality
transformations to it to obtain a non-trivial solution. In contrast,
the degrees of freedoms of the Kaluza-Klein vector and the Maxwell fields
are related to, respectively, the anti-symmetric tensor and the abelian
gauge fields in our diagonal embedding, which
enable us to read off a string solution directly
from a solution of general relativity. At the same time the new solutions
themselves obtained in this way can be ``seeds'' solutions to which
the duality transformations apply. In view of the vast
accumulation of knowledge of the solution-generating technique in general
relativity, we hope that this analogy may provide deeper insight on
the duality symmetry of string theory. Although we only consider the
colliding wave solitions in this paper, our discussion can be
straightforwardly applied to the case of the $(++)$ signature.

The plan of this paper is as follows. In Sect.2 we will briefly review
the basic facts on the dimensionally reduced Einstein gravity in the
presence of two (space-like) commuting Killing vector fields, and then
explain the geometrical picture of the Matzner-Misner and the Ehlers
symmetry. In Sect.3 we apply the inverse scattering method to the
two-dimensional effective heterotic string theory, developing the
formulation of the linear system obtained in Ref.\cite{Sen2d}. The
construction of a sample solution motivates us to consider the diagonal
embedding in the subsequent sections.
In Sect.4 we will embed the nonlinear sigma models
of the reduced Einstein and the
Einstein-Maxwell theories into the effective string theory without
and with the U(1)$^{16}$
gauge fields, respectively. We will see that either of
the integral Matzner-Misner and Ehlers SL(2,{\bf Z})
can be embedded
into a special T-duality because of the Kramer-Neugebauer involution,
and that, in particular, if the Matzner-Misner one is identified
as a T-duality, then the Ehlers one acts as the S-duality
on the four dimensional sector. In Sect.5 we will obtain some new
colliding string wave solutions by simply recasting the
Ferrari-Iba$\widetilde{\rm n}$ez infinite series of metrics, the
Nutku-Halil metric, and its electro-magnetic generalization.
Finally we summarize our results in Sect.6.

\section{Dimensionally Reduced Einstein Gravity}

\subsection{Ehlers and Matzner-Misner SL(2,R)}

As have been already stated in Introduction, the dimensionally
reduced Einstein gravity is the simplest prototype that
exhibits a non-compact ``hidden'' symmetry.
We first consider the reduction of the Einstein action
from four dimensions to three by introducing a Killing vector
field along the $x^3\equiv y$ axis
(In this section we drop the
$x^2\equiv x$ and $x^3\equiv y$ dependences while
leave $x^0$ and $x^1$ as
the coordinates of the reduced two-dimensional field theory.).

We start from the following vierbein
\beq
E_{M}^{~A}=\left[
\begin{array}{cc}
\D^{-\frac{1}{2}}e_{\bar{\mu}}^{~\bar{\alpha}}
&\D^{\frac{1}{2}} B_{\bar{\mu}}\\
0&\D^{\frac{1}{2}}
\end{array}
\right],\label{vierbein}
\eeq
where $M$ and $A$ denote the four-dimensional spacetime and Lorentz
indices, while $\bar{\mu}$ and $\bar{\alpha}$ do the corresponding
three-dimensional ones.
All the components are assumed to be independent
of the $y$ coordinate.
Up to a total derivative, the Einstein Lagrangian is reduced to
\beqa
{\cal L}&=&
\sqrt{-\mbox{det}E_M^{~~A}}R(E_M^{~~A})\label{sqrtgR}\\
&=&\sqrt{-g^{(3)}}\left[
R^{(3)}-\frac{1}{4}\D^2F_{\bar{\mu}\bar{\nu}}F^{\bar{\mu}\bar{\nu}}
-\frac{1}{2}g^{(3)\bar{\mu}\bar{\nu}}\D^{-2}
\partial_{\bar{\mu}}\D\partial_{\bar{\nu}}\D
\right],\label{L3dEinstein}
%\\&&\mbox{$\bar{\mu}$,$\bar{\nu}=0,1,2$},\nonumber
\eeqa
where
$g^{(3)}_{\bar{\mu}\bar{\nu}}
=e_{\bar{\mu}}^{~\bar{\alpha}}\eta_{\bar{\alpha}\bar{\beta}}e^{\bar{\beta}}
_{~~\bar{\nu}}$,
$F_{\bar{\mu}\bar{\nu}}=\partial_{\bar{\mu}}B_{\bar{\nu}}
-\partial_{\bar{\nu}}B_{\bar{\mu}}$
and $F^{\bar{\mu}\bar{\nu}}
=g^{(3)\bar{\mu}\bar{\sigma}}g^{(3)\bar{\nu}\bar{\lambda}}
F_{\bar{\sigma}\bar{\lambda}}$.

The Ehlers symmetry is not a symmetry of the original Einstein action,
but a symmetry of the equations of motion,
as other S-like duality symmetries are.
It is, however, known in supergravity theories
that there exist a trick to obtain an action manifestly invariant
under such symmetries.
Namely we add a Lagrange multiplier term to the Lagrangian
to treat $F^{\bar{\mu}\bar{\nu}}$ as an independent field\footnote{
Here $\epsilon^{\bar{\mu}\bar{\nu}\bar{\sigma}}$ is the ``densitized''
totally anti-symmetric tensor, while the ``undensitized'' one, which
takes the values $\pm(\sqrt{-g^{(3)}})^{-1}$, has been used in some
literature.
}
\beq
{\cal L}'={\cal L}-\frac{1}{2}B
%\cdot\sqrt{-g^{(3)}}
\epsilon^{\bar{\mu}\bar{\nu}\bar{\sigma}}\partial_{\bar{\sigma}}
F_{\bar{\mu}\bar{\nu}}.
\label{L'}
\eeq
The equation of motion of $F^{\bar{\mu}\bar{\nu}}$ reads
\beq
\D^2F^{\bar{\mu}\bar{\nu}}=
(\sqrt{-g^{(3)}})^{-1}\epsilon^{\bar{\mu}\bar{\nu}\bar{\sigma}}
\partial_{\bar{\sigma}}B.
\label{duality}
\eeq
Substituting (\ref{duality}) into (\ref{L'}), we obtain an
SL(2,R)/U(1) non-linear sigma model coupled to gravity
\beq
{\cal L}'
=\sqrt{-g^{(3)}}\left[
R^{(3)}
-\frac{1}{2}g^{(3)\bar{\mu}\bar{\nu}}\D^{-2}
(\partial_{\bar{\mu}}B\partial_{\bar{\nu}}B
+\partial_{\bar{\mu}}\D\partial_{\bar{\nu}}\D)
\right],
\eeq
which is manifestly invariant under the fractional linear
transformation
\beq
Z^{(\mbox{\scriptsize E})} \rightarrow \frac{aZ^{(\mbox{\scriptsize E})}
+b}{cZ^{(\mbox{\scriptsize E})}+d}~~~
(a,b,c,d\in\mbox{\bf R}),\label{Ehlers}
\eeq
where
\beq
Z^{(\mbox{\scriptsize E})}
=B+i\D.
\eeq
This is the Ehlers SL(2,{\bf R}) symmetry
(One may set $ad-bc=1$ without loss of generality.) \cite{Ehlers}.
$Z^{(\mbox{\scriptsize E})}$ is related to the Ernst potential $\E$ by
\beq
\E=i\overline{Z^{(\mbox{\scriptsize E})}}.
\eeq

We descend further from three to two dimensions by discarding
the $x^2\equiv x$-coordinate dependence of the fields.
Assuming the form of the vierbein as
\beq
E_M^{~~A}=\left[
\begin{array}{cc}
\Delta^{-\frac{1}{2}}\lambda\delta_{\mu}^{~\alpha}&\mbox{\large $0$}\\
\mbox{\large $0$}&\begin{array}{cc}
\Delta^{-\frac{1}{2}}\rho&\Delta^{\frac{1}{2}}B_x\\
0&\Delta^{\frac{1}{2}}
\end{array}
\end{array}\right],\label{vierbein2}
\eeq
where $\mu$ and $\nu$ stand for the two-dimensional spacetime and Lorentz
indices,
the reduced two dimensional Lagrangian is found to be
\beq
{\cal L}'=\rho\eta^{\mu\nu}
\left[
-2\partial_{\mu}\partial_{\nu}\ln\lambda
-\frac{1}{2}
\Delta^{-2}(\partial_{\mu}\Delta\partial_{\nu}\Delta
+\partial_{\mu}B\partial_{\nu}B)
\right].\label{L_E}
\eeq
The indices $\mu,\nu=0,1$ are now raised by $\eta^{\mu\nu}$.
The relation (\ref{duality}) then becomes
\beq
\D^2\partial_{\mu}B_x=
\rho\epsilon^{(2)~~\nu}_{~~~\mu}
\partial_{\nu}B,~~~\epsilon^{(2)~~\nu}_{~~~\mu}=\pm 1
\label{duality2dim}
\eeq
in two dimensions.
The twist potential $B$ and a component of the vierbein $B_x$
are related non-locally with each other by the equation
(\ref{duality2dim}).
The equations of motions of $\D$ and $B$ are compactly
described by the Ernst equation \cite{Ernst}
\beq
\rho\partial_{\mu}\E\partial^{\mu}\E
=\D\partial_{\mu}(\rho\partial^{\mu}\E).\label{Ernsteq}
\eeq

Another way to obtain a two-dimensional model is to reduce the
dimensions from four to two directly. Substituting the parameterization
(\ref{vierbein2}) in (\ref{sqrtgR}), we find
\beq
{\cal L}=\rho\eta^{\mu\nu}
\left[
-2
\partial_{\mu}\partial_{\nu}
\ln(\lambda\D^{-\frac{1}{2}}\rho^{\frac{1}{4}})
-\frac{\D^2}{2\rho^2}\left\{
\partial_{\mu}\left(\frac{\rho}{\D}\right)
\partial_{\nu}\left(\frac{\rho}{\D}\right)
+\partial_{\mu}B_x\partial_{\nu}B_x
\right\}
\right].\label{L_MM}
\eeq
Defining
\beq
Z^{(\mbox{\scriptsize MM})}=B_x+i\frac{\rho}{\D}
\eeq
in this case, the
Lagrangian (\ref{L_MM}) is invariant under the Matzner-Misner
SL(2,R) transformation
\beq
Z^{(\mbox{\scriptsize MM})} \rightarrow
\frac{aZ^{(\mbox{\scriptsize MM})}+b}{cZ^{(\mbox{\scriptsize MM})}+d}~~~
(a,b,c,d\in\mbox{\bf R};~ad-bc=1).\label{MatznerMisner}
\eeq
The two SL(2,{\bf R}) groups (\ref{Ehlers}) and (\ref{MatznerMisner})
generate the Geroch group.
It is well-known that they are nontrivially entangled with each other
to constitute an infinite-dimensional sl(2,{\bf R}) Kac-Moody algebra
\cite{Julia}. The replacement of the variables
\beq
B\leftrightarrow B_x,~~\D\leftrightarrow\frac{\rho}{\D},
{}~~\lambda\leftrightarrow
\lambda\D^{-\frac{1}{2}}\rho^{\frac{1}{4}},~~
\rho\leftrightarrow\rho, \label{KN}
\eeq
which makes the Lagrangians (\ref{L_E}) and (\ref{L_MM}) identical,
is called the Kramer-Neugebauer involution \cite{KN}.
Let $\E^{(\mbox{\scriptsize MM})}$ be the image of $\E$ by (\ref{KN}),
then obviously $\E^{(\mbox{\scriptsize MM})}$ also
satisfies the ``Ernst equation''
\beq
\rho\partial_{\mu}\E^{(\mbox{\scriptsize MM})}\partial^{\mu}
\E^{(\mbox{\scriptsize MM})}
=\frac{\rho}{\D}\partial_{\mu}(\rho\partial^{\mu}
\E^{(\mbox{\scriptsize MM})}).
\eeq

\subsection{Integral Matzner-Misner SL(2,Z) as
the Modular Transformation of a Torus}\label{2.2}

It is not an accident that the Teichm\"{u}ller space of a torus
($=$ the upper-half plane) appears as the target space of the
sigma model of the dimensionally reduced Lagrangian (\ref{L_MM}).
To clarify this point, let us assume that the two commuting Killing
vector fields are tangent to a two-torus. In other words we consider
the ``toroidal compactification'' of Einstein gravity. The resulting
action of the reduced theory depends only on the two coordinates
$x^{\mu}$, $\mu=0,1$, and each point $x^{\mu}$ associates a flat torus
parameterized by the coordinates $x^m$, $m=2,3$.
The constant metric of this torus is given by the zweibein
\beq
E_m^{~a}=\left[
\begin{array}{cc}
\Delta^{-\frac{1}{2}}\rho&\Delta^{\frac{1}{2}}B_x\\
0&\Delta^{\frac{1}{2}}
\end{array}\right]. \label{zweibein1}
\eeq
The line element reads
\beq
%ds^2&=&-\Delta^{-1}\lambda^2\eta^{\mu\nu}dx^{\mu}dx^{\nu}\n
dL^2=\Delta\left[
(dy+B_xdx)^2+(\frac{\rho}{\Delta}dx)^2
\right]~~~(x^2\equiv x, x^3\equiv y).\label{dL1}
\eeq

It is well-known that each point $z=\tau$ in the upper-half complex
plane represents an inequivalent complex structure of a torus.
For the above metric (\ref{dL1}) it turns out that
the modular parameter $\tau$ coincides with
$Z^{(\mbox{\scriptsize MM})}$. Indeed, this can be checked as follows:
$\tau$ $(\mbox{Im}\tau>0)$ stands
for a constant metric of a torus obtained as a lattice
on the complex plane, whose periods are 1 and $\tau$. Mapping this
rectangle to a square $0\leq x\leq 1$, $0\leq y\leq 1$,
the metric becomes
\beq
dL^2=(dy+\mbox{Re}\tau dx)^2+(\mbox{Im}\tau dx)^2. \label{dL2}
\eeq
Comparing (\ref{dL1}) with (\ref{dL2}), we find that the conformal
structure of the metric (\ref{dL1}) is represented by the modular
parameter
\beq
\tau=B_x+i\frac{\rho}{\D}
\eeq
(Note that all $\D$, $\rho$ and $B_x$ are constant with respect to
the coodinates of the torus $x,y$.).

It is now easy to see the meaning of the Matzner-Misner symmetry;
this is nothing but the invariance of the system under the change
of conformal structure of the compactified torus. Of course, an
infinitesimal change can be a symmetry because we consider a
Kaluza-Klein theory, i.e. only the constant modes on the compactified
space. Therefore, if we consider the full theory in which all the
massive excitations are included, then we may only have the discrete
modular group symmetry
\beq
Z^{(\mbox{\scriptsize MM})}
\rightarrow \frac{aZ^{(\mbox{\scriptsize MM})}+b}
{cZ^{(\mbox{\scriptsize MM})}+d}~~~
(a,b,c,d\in\mbox{\bf Z};~ad-bc=1).\label{integralMatznerMisner}
\eeq
We will see in the next subsection that this is a special T duality
if embedded in string theory.

What is the meaning of the Ehlers symmetry, then? We can see this
by using the Kramer-Neugebauer involution, by which
the zweibein (\ref{zweibein1}) is mapped to
\beq
E_m^{~a}=\left(\frac{\rho}{\D}\right)^{\frac{1}{2}}\left[
\begin{array}{cc}
\Delta&B\\
0&1
\end{array}\right]. \label{zweibein2}
\eeq
The line element reads
\beq
dL^2=\frac{\rho}{\D}\left[
(dy+Bdx)^2+(\Delta dx)^2
\right].
\eeq
Again it is easy to see that $Z^{(\mbox{\scriptsize E})}$ coincides
to the modular parameter of the metric (\ref{zweibein2}), but
in this case this metric is the one on a ``fictitious'' torus;
one of the component $B$ is related
non-locally  to a component $B_x$ of a ``real'' torus. The Ernst
potential $\E$ is then essentially a modular parameter of this
torus\footnote{In the Ashtekar formulation of the
dimensionally reduced gravity, the Ashtekar connections
corresponding to the compactified direction were shown to be
modular forms \cite{Mizoguchi}.}.
The Ehlers integral fractional linear transformation
\beq
Z^{(\mbox{\scriptsize E})} \rightarrow
\frac{aZ^{(\mbox{\scriptsize E})}
+b}{cZ^{(\mbox{\scriptsize E})}+d}~~~
(a,b,c,d\in\mbox{\bf Z}),\label{integralEhlers}
\eeq
is then the modular transformation of this torus.
In particular, the modular S transformation\footnote{
Of course this confusing name has nothing to do with the fact,
which we will show later, that the Ehlers SL(2,{\bf Z}) acts
as the S duality.
}
\beq
Z^{(\mbox{\scriptsize E})} \rightarrow
-\frac{
1}{Z^{(\mbox{\scriptsize E})}},\label{EhlersSmod}
\eeq
which is a special element
of the Ehlers SL(2,{\bf Z}) group,
is simply referred to as the ``Ehlers transformation''
in the literature.

\section{Linear System and Inverse Scattering Method}\label{Inverse}
\subsection{Linear System}
In this section we first review the construction of the linear
system of the two-dimensional heterotic string theory \cite{Sen2d},
and then use this formulation to obtain a classical solution by
means of the inverse scattering method.
We will closely follow the Ref.\cite{Nicolai}.

It has been shown \cite{MaharanaSchwarz} that the bosonic sector of the
ten-dimensional effective low-energy action of the heterotic string
theory
\beqa
S&=&\int d^{10}z\sqrt{-G^{(10)}}e^{-\Phi^{(10)}}
\left(R^{(10)}+G^{(10)MN}\partial_M\Phi^{(10)}\partial_N\Phi^{(10)}
\right.\n
&&\left.~~~~~~~~-\frac{1}{12}H_{MNP}^{(10)}H^{(10)MNP}
-\frac{1}{4}F_{MN}^{(10)I}F^{(10)IMN}\right),\label{tendimaction}\\
F_{MN}^{(10)I}&=&\partial_MA_N^{(10)I}-\partial_NA_M^{(10)I},\\
H_{MNP}^{(10)}&=&\left(
\partial_MB_{NP}^{(10)}-\frac{1}{2}A_M^{(10)I}F_{NP}^{(10)I}
\right)+(\mbox{cyclic permutations}),\n
&&(M,N,P=0,\ldots,9)\nonumber
\eeqa
is
reduced to the two dimensional action
\beq
S=\int d^2x\sqrt{-G}e^{-\Phi}
\left[
R_{G}+G^{\mu\nu}\partial_{\mu}\Phi\partial_{\nu}\Phi
+\frac{1}{8}G^{\mu\nu}{\rm Tr}(\partial_{\mu}ML\partial_{\nu}ML)
\right]\label{action}
\eeq
if compactified on an eight-torus, where
the ten-dimensional metric $G^{(10)}_{MN}$,
anti-symmetric tensor field $B^{(10)}_{MN}$,
U(1)$^{16}$ gauge field $A_M^{(10)I}$, and the dilaton
field $\Phi^{(10)}$ are assumed to take the forms
\beqa
&&G_{MN}^{(10)}=\left[\begin{array}{cc}
G_{\mu\nu}&0\\0&\widehat{G}_{mn}
\end{array}\right],~
B_{MN}^{(10)}=\left[\begin{array}{cc}
B_{\mu\nu}=0&0\\0&\widehat{B}_{mn}
\end{array}\right],~\n
&&A_{M}^{(10)I}=\left[\begin{array}{c}
A_{\mu}^I=0\\
\widehat{A}_m^I
\end{array}\right],~e^{-\Phi^{(10)}}
=(\mbox{det}\widehat{G})^{-\frac{1}{2}}e^{-\Phi}.\label{strbkgrd1}
\eeqa
In (\ref{strbkgrd1}) some fields without physical degrees of freedom
in two dimensions have been set to zero.
$x^{\mu}$ $(\mu=0,1)$ and $y^m$ $(m=1,\ldots,8)$
are the coordinates of the two dimensional space-time and
the eight-dimensional compactified torus, respectively.
The $32\times 32$ matrices $M$ and $L$ are given by
\beq
M=\left[
\begin{array}{ccc}
\widehat{G}^{-1}&
\widehat{G}^{-1}(\widehat{B}+\widehat{C})&\widehat{G}^{-1}\widehat{A}\\
(-\widehat{B}+\widehat{C})\widehat{G}^{-1}&
(\widehat{G}-\widehat{B}+\widehat{C})
\widehat{G}^{-1}(\widehat{G}+\widehat{B}+\widehat{C})&
(\widehat{G}-\widehat{B}+\widehat{C})\widehat{G}^{-1}\widehat{A}\\
\widehat{A}^T\widehat{G}^{-1}&
\widehat{A}^T\widehat{G}^{-1}(\widehat{G}+\widehat{B}+\widehat{C})&
I_{16}+\widehat{A}^T\widehat{G}^{-1}\widehat{A}
\end{array}
\right]\label{M}
\eeq
with $\widehat{C}=\frac{1}{2}\widehat{A}\widehat{A}^T$, and
\beq
L=\left[
\begin{array}{ccc}
&I_8&\\I_8&&\\&&-I_{16}
\end{array}
\right].\label{L}
\eeq
They satisfy
$M^T=M$, $MLM^T=L$.
We further define
\beq
U_0=\left[
\begin{array}{ccc}
\frac{1}{\sqrt{2}}I_8&-\frac{1}{\sqrt{2}}I_8&\\
\frac{1}{\sqrt{2}}I_8&\frac{1}{\sqrt{2}}I_8&\\
&&I_{16}
\end{array}\right]
\eeq
and
\beq
U_0^TMU_0\equiv M'.
\eeq
$M'$ is again a symmetric matrix $M'^T=M'$, and
\beq
U_0^TLU_0=\left[
\begin{array}{ccc}
I_8&&\\
&-I_8&\\
&&-I_{16}
\end{array}
\right]\equiv I_{8,24}.
\eeq
$M'$ belongs to O(8,24), preserving the bilinear form defined by
$I_{8,24}$ invariant
\beq
M'^TI_{8,24}M'=I_{8,24}.
\eeq
The Lie algebra o(8,24) consists of the matrices
\beq
\left\{
\left.\left[\begin{array}{cc}X_1&X_2\\X_2^T&X_3\end{array}\right]\right|
{}~X_1^T=-X_1,~~X_3^T=-X_3
\right\}.
\eeq
$X_2$ is an arbitrary real $8\times 24$ matrix.
The symmetric-space automorphism $\tau$ of the coset
O(8,24)/O(8)$\times$O(24) is given by \cite{symspace}
\beq
\tau(X)=I_{8,24}XI_{8,24}~~~(X\in\mbox{\rm o(8,24)}).\label{tau}
\eeq
o(8,24) is decomposed into eigenspaces of $\tau$ as
\beqa
&&\mbox{\rm o(8,24)}=\mbox{\bf K}\oplus\mbox{\bf H},
\label{decomp}\\
&&\mbox{\bf K}=
\left\{
\left.\left[\begin{array}{cc}0&X_2\\X_2^T&0\end{array}\right]\right|
{}~X_2:\mbox{arbitrary real $8\times 24$ matrix}
\right\},\\
&&\mbox{\bf H}=
\left\{
\left.\left[\begin{array}{cc}X_1&0\\0&~X_3\end{array}\right]\right|
{}~X_1^T=-X_1,~X_3^T=-X_3
\right\},
\eeqa
so that $\tau({\mbox{\bf K}})=-\mbox{\bf K}$,
$\tau({\mbox{\bf H}})=\mbox{\bf H}$.
$\mbox{\bf H}$ is the Lie algebra of the denominator
O(8)$\times$O(24) of the coset. Since
$X^T=X$ if $X\in\mbox{\bf K}$, and
$X^T=-X$ if $X\in\mbox{\bf H}$, (\ref{tau}) is equivalent to
\beq
\tau(g)=(g^T)^{-1}~~~(g\in\mbox{\rm O(8,24)}).\label{tau2}
\eeq

It is known that $M'$ can be written as
\beq
M'={\cal V}{\cal V}^T,
\eeq
where $\cV\in\mbox{\rm O(8,24)}$ is given by
\cite{MaharanaSchwarz}
\beq
{\cal V}=U_0^TVU_0,~~
V=\left[\begin{array}{ccc}
\hE^{-1}&0&0\\
(-\widehat{B}+\widehat{C})\hE^{-1}&\hE^T&\widehat{A}\\
\widehat{A}^T\hE^{-1}&0&I_{16}
\end{array}\right].
\label{Vhetero}
\eeq
$\hE$ is an ``achtbein'' (vielbein with eight indices) of
$\widehat{G}$, satisfying
$\widehat{G}=\hE^T\hE$.
Using the decomposition (\ref{decomp})
we write
\beq
{\cal V}^{-1}\partial_{\mu}{\cal V}=P_{\mu}+Q_{\mu},
{}~~P_{\mu}\in\mbox{\bf K},
{}~~Q_{\mu}\in\mbox{\bf H},\label{P+Q}
\eeq
then
\beqa
&&\mbox{\rm Tr}(\partial_{\mu}ML\partial_{\nu}ML)\n
&=&-\mbox{\rm Tr}(M^{-1}\partial_{\mu}M\cdot M^{-1}\partial_{\nu}M)\n
&=&-\mbox{\rm Tr}({M'}^{-1}\partial_{\mu}{M'}\cdot {M'}^{-1}
\partial_{\nu}{M'})\n
&=&-\mbox{\rm Tr}\left[
\left(\cV^{-1}\partial_{\mu}\cV+(\cV^{-1}\partial_{\mu}\cV)^T\right)
\cdot
\left(\cV^{-1}\partial_{\nu}\cV+(\cV^{-1}\partial_{\nu}\cV)^T\right)
\right]\n
&=&-4\mbox{\rm Tr}P_{\mu}P_{\nu}.\label{TrPP}
\eeqa
In the last line we used $P_{\mu}^T=P_{\mu}$ and
$Q_{\mu}^T=-Q_{\mu}$.
The O(8)$\times$O(24) gauge transformation $\cV\mapsto\cV h(x)$,
$h(x)\in\mbox{O(8)$\times$O(24)}$, which does not change $M'$,
acts on $P_{\mu}$ and $Q_{\mu}$ as
\beq
P_{\mu}\mapsto h^{-1}P_{\mu}h,~~
Q_{\mu}\mapsto h^{-1}Q_{\mu}h+h^{-1}\partial_{\mu}h.
\eeq
Hence (\ref{TrPP}) is manifestly gauge invariant.
The dilaton kinetic term can be absorbed into the $R_G$ term
by a rescaling
\beq
G_{\mu\nu}=e^{\Phi}g_{\mu\nu}, \label{Gandg}
\eeq
so that
the action becomes
\beq
S=\int d^2xe^{-\Phi}
\left[
\sqrt{-g}R_{g}
-\frac{1}{2}\eta^{\mu\nu}{\rm Tr}P_{\mu}P_{\nu}
\right],\label{action2}
\eeq
where we have adopted the conformal gauge
\beq
g_{\mu\nu}=\lambda^2\eta_{\mu\nu}.
\label{conformalgauge}
\eeq
This is precisely the same Lagrangian resulting from the reduction
of the Einstein Lagrangian from four dimensions to two,
except the difference of the compactified sector.
This observation is a clue to construct a mapping
from solutions of general relativity to those of string theory.
The independent equations of motion arising from the action
(\ref{action2}) are
\beqa &&\partial_+\partial_-e^{-\Phi}=0,\label{EM1}\\
&&\lambda^{-1}\partial_{\pm}\lambda\cdot\rho\partial_{\pm}\rho
=\frac{1}{2}\rho^{-1}\partial_{\pm}^2\rho
+\frac{1}{4}\mbox{\rm Tr}P_{\pm}P_{\pm},\label{EM2}\\
&&D_+(\rho P_-)+D_-(\rho P_+)=0,\label{EM3}
\eeqa where we switched to the light-cone coordinate, and
\beq
D_{\mu}P_{\nu}\equiv\partial_{\mu}P_{\nu}+{[}Q_{_{\mu}},~P_{\nu}{]}.
\eeq

We next introduce the spectral-parameter dependent matrix $\hcV$
\cite{xdept,BM} defined by
\beq
\hcV^{-1}\partial_{\mu}\hcV=Q_{\mu}+\frac{1+t^2}{1-t^2}P_{\mu}
+\frac{2t}{1-t^2}\epsilon_{\mu\nu}P^{\nu}.
\eeq
The zero-curvature condition
\beq
\partial_{\mu}(\hcV^{-1}\partial_{\nu}\hcV)
-\partial_{\nu}(\hcV^{-1}\partial_{\mu}\hcV)
+\left[
\hcV^{-1}\partial_{\mu}\hcV,~\hcV^{-1}\partial_{\nu}\hcV
\right]=0
\eeq
is equivalent to the relevant equation of motion (\ref{EM3})
and the ``flat-space'' zero-curvature condition (for $\cV=\hcV(t=0)$)
\cite{Pohlmeyer},
if the spectral parameter $t$ satisfies \cite{xdept,BM,Nicolai}
\beq
\partial_{\pm}\ln\rho=\frac{1\mp t}{1\pm t}\partial_{\pm}\ln t.
\eeq
This equation can be integrated to give
\beq
w=\frac{1}{4}(\rho_+(x^+)+\rho(x^-))\left(
t+\frac{1}{t}
\right)-\frac{1}{2}(\rho_+(x^+)-\rho(x^-))
\eeq
with some real constant $w$, where we solved the equation of motion of
the dilaton $\Phi$ to decompose it into left- and right-moving components
\beq
e^{-\Phi}\equiv\rho\equiv\rho_+(x^+)+\rho(x^-).
\eeq

\subsection{Inverse Scattering Method}
To demonstrate how this formulation works, let us construct
the simplest colliding string wave solution.
In our O(8,24)/O(8)$\times$O(24) case, the generalized symmetric-space
automorphism $\tau^{\infty}$ turns out to be
\beqa
\tau^{\infty}\left(\hcV(t)\right)&\equiv& \tau(\hcV(t^{-1}))\n
&=&I_{8,24}\hcV(t^{-1})I_{8,24}\n
&=&\left[(\hcV(t^{-1}))^T
\right]^{-1}.
\label{tauinfty}
\eeqa
This has the following properties
\beq
P_{\mu}\rightarrow -P_{\mu},~~Q_{\mu}\rightarrow Q_{\mu},~~
t\rightarrow t^{-1},
\eeq
and
\beq
\tau^{\infty}\left(\hcV^{-1}\partial_{\mu}\hcV\right)
=\hcV^{-1}\partial_{\mu}\hcV.\label{tauinv}
\eeq
Owing to the invariance (\ref{tauinv}), the monodromy matrix
\beq
{\cal M}\equiv\hcV\tau^{\infty}(\hcV^{-1}) \label{factorizedM}
\eeq
does not depend on $x^{\mu}$. Hence one may
construct a solution by finding ${\cal M}$ that factorizes
into $\hcV$ and $\tau^{\infty}(\hcV^{-1})$,
and then calculating the conformal factor $\lambda^2$ for such
$\hcV$.

Motivated by the example discussed in Ref.\cite{Nicolai},
let us take the following monodromy matrix
\beq
{\cal M}=U_0^T\left[
\begin{array}{ccc}
-\frac{w-\frac{1}{2}}{w+{\frac{1}{2}}}I_8&&\\
&-\frac{w+\frac{1}{2}}{w-{\frac{1}{2}}}I_8&\\
&&\mbox{\Large $0$}
\end{array}
\right]U_0.\label{calM}
\eeq
Fixing the conformal invariance by choosing
\beq
\rho_+=\frac{1}{2}-(x^+)^2\equiv\frac{1}{2}-u^2,~~~
\rho_-=\frac{1}{2}-(x^-)^2\equiv\frac{1}{2}-v^2,
\label{uv}
\eeq
it is straightforward to check that this choice of ${\cal M}$
allows the factorization (\ref{factorizedM}) as follows:
\beq
\hcV=U_0^T\left[
\begin{array}{ccc}
-\sqrt{-\frac{t_2}{t_1}}\frac{t-t_1}{t-t_2}I_8&&\\
&-\sqrt{-\frac{t_1}{t_2}}\frac{t-t_2}{t-t_1}I_8&\\
&&\mbox{\Large $0$}
\end{array}
\right]U_0,
\eeq
\beq
t_1=\frac{\sqrt{1-u^2}-v}{\sqrt{1-u^2}+v},~~~
t_2=\frac{u+\sqrt{1-v^2}}{u-\sqrt{1-v^2}}.
\eeq
($-\frac{t_2}{t_1}\in${\bf R}, $>0$ in the interaction region
$u,v>0$ and $u^2+v^2<1$.)
Note that the ${\cal M}$ matrix (\ref{calM}) is factorized in the
same way as is done in the general relativity case, because of
the formal reminiscence of $\tau^{\infty}$ (\ref{tauinfty}).
{}From (\ref{Vhetero}) we find that this choice of ${\cal M}$
corresponds to the configurations
\beq
\hG_{mn}=-\frac{t_2}{t_1}\delta_{mn},~~~\hB_{mn}=0,~~~\hA_{m}=0.
\eeq

The conformal factor $\lambda$ can be calculated by integrating the
equations (\ref{EM2}), which now read
\beq
\lambda^{-1}\partial_{\pm}\lambda\cdot\rho\partial_{\pm}\rho
=\frac{1}{2}\rho^{-1}\partial_{\pm}^2\rho
+\left(\partial_{\pm}\ln\frac{-t_2}{t_1}
\right)^2.
\label{EM2'}
\eeq
Without any detailed calculation, we can immediately find the expression
of $\lambda$ if we notice that this equation is completely the same
as the one that appears in the construction of a
Ferrari-Iba$\widetilde{\rm n}$ez metric in Ref. \cite{Nicolai},
except the extra factor 4 in the second term of the left hand side
in (\ref{EM2'}).
The result is
\beq
\lambda^2=\mbox{const.}\times
uv\left(
\frac{(1-t_1t_2)^2}{(1-t_1^2)(1-t_2^2)}
\right)^4.
\eeq

We have seen in this example that the problem to
solve the string equations of motion can be deduced to almost the same
problem arising in general relativity, if we assume some special
form of the string backgrounds. Of course one could obtain more
non-trivial solutions if one starts from more complicated
monodromy matrices. We will not do this, but will pursue the similarity
between the general relativity and string theory in the subsequent
sections.

\section{Embedding General Relativity into String Theory}\label{Embedding}
\subsection{Embedding Einstein Theory into O(8,8) Theory}
In this section we will identify
more general string backgrounds such that they satisfy the
(electro-magnetic) Ernst equation.
We first switch off the $\widehat{A}_{\mu}$ fields, i.e.
work with the O(8,8)/ O(8)$\times$O(8) theory.
We may hence consider only the upper-left 16$\times$16 matrices.
The $V$ matrix is then reduced to
\beq
V=\left[\begin{array}{cc}
\hE^{-1}&0\\
-\widehat{B}\hE^{-1}&\hE^T
\end{array}\right].
\eeq
%and hence
%\beq
%(V^T)^{-1}=\left[\begin{array}{cc}
%\hE^T&-\hB\hE^{-1}\\
%0&\hE^{-1}
%\end{array}\right].
%\eeq
%The latter is already in a suggestive form similar to the
%parameterization used in the SL(2,{\bf R})/U(1) sigma model
%\cite{Nicolai}.
%\beqa
%V^T\partial_{\mu}((V^T)^{-1})
%&=&U\left[
%\cV^T\partial_{\mu}((\cV^T)^{-1})
%\right]U^T\n
%&=&U\left[
%\partial_{\mu}((\cV)^{-1})
%\cV\right]^TU^T\n
%&=&U(-P_{\mu}+Q_{\mu})U^T,
%\eeqa
Making use of the relations (\ref{Vhetero})(\ref{P+Q}),
we obtain
\beqa
&&P_{\mu}\n
&=&
\frac{1}{2}\left[
\begin{array}{cc}
0&\mbox{\scriptsize
$\hE^{-1}\partial{\mu}\hE+\partial{\mu}\hE\cdot\hE^{-1}
+\hE^{-1}\partial_{\mu}\hB\cdot\hE^{-1}
$}
\\
\mbox{\scriptsize
$\hE^{-1}\partial{\mu}\hE+\partial{\mu}\hE\cdot\hE^{-1}
-\hE^{-1}\partial_{\mu}\hB\cdot\hE^{-1}$}
&0
\end{array}
\right],\n
&&\label{P}\\
&&Q_{\mu}\n
&=&
\frac{1}{2}\left[
\begin{array}{cc}
\mbox{\scriptsize
$\hE^{-1}\partial{\mu}\hE-\partial{\mu}\hE\cdot\hE^{-1}
-\hE^{-1}\partial_{\mu}\hB\cdot\hE^{-1}
$}&0
\\0&
\mbox{\scriptsize
$\hE^{-1}\partial{\mu}\hE-\partial{\mu}\hE\cdot\hE^{-1}
+\hE^{-1}\partial_{\mu}\hB\cdot\hE^{-1}$}
\end{array}
\right],\n
&&\label{Q}
\eeqa
where we have taken $\hE$ to be symmetric by using the local
Lorentz rotation in the compactified sector.
The second term in $[\cdots]$ of (\ref{action2}) then reads
\beqa
&&\eta^{\mu\nu}\mbox{\rm Tr}P_{\mu}P_{\nu}\n
&=&\eta^{\mu\nu}
\frac{1}{2}\mbox{\rm Tr}
\left\{
(\hE^{-1}\partial_{\mu}\hE+\partial_{\mu}\hE\cdot\hE^{-1})
(\hE^{-1}\partial_{\nu}\hE+\partial_{\nu}\hE\cdot\hE^{-1})
\right.\n
&&~~~~~\left.
-\hE^{-1}\partial_{\mu}\hB\hE^{-1}\hE^{-1}\partial_{\nu}\hB\hE^{-1}
\right\},\label{TrPP2}
\eeqa
where the trace in the right hand side is the one for $8\times 8$
matrices. Let us now assume that $\hE$ and $\hB$ are in the forms
\beq
\hE=\Delta^{\frac{1}{2}}X,~~~
\hB=BY,\label{hEhB}
\eeq
for some scalar fields $\Delta$ and $B$, where $X$ is an arbitrary
constant $8\times 8$ matrix satisfying $X^2=I_8$, and
$Y$ is an arbitrary constant $8\times 8$ antisymmetric matrix.
We normalize $Y$ as
\beq
\mbox{\rm Tr}Y^2=-8
\eeq
for convenience. For example let us take
\beq
X=I_8,~~~Y=\left[
\begin{array}{cc}&I_4\\-I_4&\end{array}
\right].\label{XY}
\eeq
Substituting (\ref{hEhB}) and (\ref{XY}) into (\ref{P}), we then
find
\beq
P_{\mu}=\frac{1}{2}\left[\begin{array}{cc}
\mbox{\Large 0}&
\begin{array}{cc}
\Delta^{-1}\partial_{\mu}\Delta I_4&
\Delta^{-1}\partial_{\mu}B I_4\\
-\Delta^{-1}\partial_{\mu}B I_4&
\Delta^{-1}\partial_{\mu}\Delta I_4
\end{array}\\
\begin{array}{cc}
\Delta^{-1}\partial_{\mu}\Delta I_4&
\Delta^{-1}\partial_{\mu}B I_4\\
-\Delta^{-1}\partial_{\mu}B I_4&
\Delta^{-1}\partial_{\mu}\Delta I_4
\end{array}&\mbox{\Large 0}
\end{array}
\right].
\eeq
(\ref{TrPP2}) reads
\beq
\eta^{\mu\nu}\mbox{\rm Tr}P_{\mu}P_{\nu}
=\frac{1}{2}\cdot 8\cdot\eta^{\mu\nu}
\Delta^{-2}(\partial_{\mu}\Delta\partial_{\nu}\Delta
+\partial_{\mu}B\partial_{\nu}B).
\eeq
The total action (\ref{action2}) now has the form
\beq
S=\int d^2xe^{-\Phi}
\left[
\sqrt{-g}R_{g}
-2\eta^{\mu\nu}
\Delta^{-2}(\partial_{\mu}\Delta\partial_{\nu}\Delta
+\partial_{\mu}B\partial_{\nu}B).
\right].\label{action3}
\eeq
This action is ``almost'' identical to the SL(2,{\bf R})/U(1)
sigma model that arises in the reduction of the Einstein action
from four dimensions to two. The only difference is the
coupling constant of the sigma model kinetic terms,
which cannot be absorbed by rescaling of fields.

This fact implies the following significant consequences. First, we
observe that the both actions lead to the same equation of motion of
$\Delta$ and $B$ fields --- the Ernst equation. This means that we may
construct solutions of the equation of motion of $\hG$ and $\hB$ fields
in a low-energy string theory from the known solutions of the Ernst
equation in general relativity. Second, the conformal factor $\lambda$
is determined by integrating the ``Virasoro conditions'' (\ref{EM2})
(the equations of motion of the missing component of $G_{\mu\nu}$),
which now reads
\beq
\lambda^{-1}\partial_{\pm}\lambda\cdot\rho\partial_{\pm}\rho
=\frac{1}{2}\rho^{-1}\partial_{\pm}^2\rho
+1\cdot
\frac{(\partial_{\pm}\Delta)^2+(\partial_{\pm}B)^2}{\Delta^2}.
\eeq
It is instructive to compare these expression with the corresponding
equations of motion in general relativity, which are given by
(See \cite{Nicolai})
\beq
\mbox{\rm (General relativity)}~~~~~
\lambda_{\mbox{\scriptsize gr}}^{-1}
\partial_{\pm}\lambda_{\mbox{\scriptsize gr}}
\cdot\rho\partial_{\pm}\rho
=\frac{1}{2}\rho^{-1}\partial_{\pm}^2\rho
+\frac{1}{4}\cdot
\frac{(\partial_{\pm}\Delta)^2+(\partial_{\pm}B)^2}{\Delta^2}.
\eeq
The difference by a factor 4 comes from the coefficient $1/2$
of the second term of (\ref{action2}), which is 1 in the case of
general relativity, times 8: the dimensions of the compactified space.
Therefore, the conformal factor of the string is just
the one of the general relativity to the 4, except the common
explicitly $\rho$-dependent factor. To be more precise, they are
related by
\beq
\frac{\lambda^2}{uv}
=\left(\frac{\lambda^2_{\mbox{\scriptsize gr}}}{uv}\right)^4.
\label{lambdarelation}
\eeq
in terms of the $(u,v)$ coordinates
(\ref{uv}).

\subsection{Embedding Einstein-Maxwell Theory into
Heterotic String Theory}
In the last subsection we saw that the nonlinear sigma model
associated with the reduced pure Einstein theory can be embedded
into the dimensionally reduced O(8,8) theory, where all the
gauge fields $\hA_{\mu}$  are set to be zero, as
the Ernst potential contains only two degrees of freedom.
We next switch on the $\hA_{\mu}$ fields, and will show that the
nonlinear sigma model associated with the Einstein-Maxwell theory
can be embedded into that of the heterotic string with such field
configurations. We now consider the following forms of the fields
\beq
\hE=\Delta^{\frac{1}{2}}I_8,~
\hB=BY,~
\hA=\left[Z~~W
\right]. \label{hEhBhA}
\eeq
Here $Z$ and $W$ are $8\times 8$ matrices that depends on some
two scalar fields. After some calculation we find
\beqa
&&P_{\mu}\n
&=&
\left[\begin{array}{cccc}
0&\mbox{\scriptsize $\begin{array}{c}\frac{1}{2}\D^{-1}\partial_{\mu}\D
+\frac{1}{2}\D^{-1}\partial_{\mu}BY
\\
-\frac{1}{4}\D^{-1}\left(
\partial_{\mu}Z\cdot Z^T-Z\partial_{\mu}Z^T\right.\\
\left.
+\partial_{\mu}W\cdot W^T-W\partial_{\mu}W^T
\right)\end{array}$}
&\frac{1}{\sqrt{2}}\D^{-\frac{1}{2}}\partial_{\mu}Z
&\frac{1}{\sqrt{2}}\D^{-\frac{1}{2}}\partial_{\mu}W\\
\mbox{\scriptsize $\begin{array}{c}\frac{1}{2}\D^{-1}\partial_{\mu}\D
-\frac{1}{2}\D^{-1}\partial_{\mu}BY
\\
+\frac{1}{4}\D^{-1}\left(
\partial_{\mu}Z\cdot Z^T-Z\partial_{\mu}Z^T\right.\\
\left.
+\partial_{\mu}W\cdot W^T-W\partial_{\mu}W^T
\right)\end{array}$}
&0&0&0\\
\frac{1}{\sqrt{2}}\D^{-\frac{1}{2}}\partial_{\mu}Z^T&0&0&0\\
\frac{1}{\sqrt{2}}\D^{-\frac{1}{2}}\partial_{\mu}Z^T&0&0&0
\end{array}
\right],\n
&&\\
&&Q_{\mu}\n
&=&\left[\begin{array}{cccc}
\mbox{\scriptsize $\begin{array}{c}
-\frac{1}{2}\D^{-1}\partial_{\mu}BY
\\
+\frac{1}{4}\D^{-1}\left(
\partial_{\mu}Z\cdot Z^T-Z\partial_{\mu}Z^T\right.\\
\left.
+\partial_{\mu}W\cdot W^T-W\partial_{\mu}W^T
\right)\end{array}$}
&0&0&0\\
0&\mbox{\scriptsize $\begin{array}{c}
\frac{1}{2}\D^{-1}\partial_{\mu}BY
\\
-\frac{1}{4}\D^{-1}\left(
\partial_{\mu}Z\cdot Z^T-Z\partial_{\mu}Z^T\right.\\
\left.
+\partial_{\mu}W\cdot W^T-W\partial_{\mu}W^T
\right)\end{array}$}
&\frac{1}{\sqrt{2}}\D^{-\frac{1}{2}}\partial_{\mu}Z
&\frac{1}{\sqrt{2}}\D^{-\frac{1}{2}}\partial_{\mu}W\\
0&-\frac{1}{\sqrt{2}}\D^{-\frac{1}{2}}\partial_{\mu}Z^T&0&0\\
0&-\frac{1}{\sqrt{2}}\D^{-\frac{1}{2}}\partial_{\mu}W^T&0&0
\end{array}\right],\n
\eeqa
where we omitted $I_8$. Let us now assume
\beq
Z=W=HI_8+EY \label{Z=W}
\eeq
for some scalar fields $E$ and $H$. This special parameterization
%(\ref{Z=W})
yields
\beqa
\eta^{\mu\nu}\mbox{\rm Tr}P_{\mu}P_{\nu}
&=&4\eta^{\mu\nu}\left[
\D^{-2}\partial_{\mu}\D\partial_{\nu}\D\right.\n
&&+\D^{-2}\left(
\partial_{\mu}B+2(E\partial_{\mu}H-H\partial_{\mu}E)\right)
\left(
\partial_{\nu}B+2(E\partial_{\nu}H-H\partial_{\nu}E)\right)\n
&&\left.+4\D^{-1}(\partial_{\mu}E\partial_{\nu}E+
\partial_{\mu}H\partial_{\nu}H)
\right].
\eeqa
This is the SU(2,1)/S(U(2)$\times$U(1)) nonlinear
sigma model, which is known to arise as a result of the dimensional
reduction of the Einstein-Maxwell theory \cite{Carter,Mazur}.
The particular configuration (\ref{hEhBhA}) with (\ref{Z=W})
thus satisfies the electro-magnetic Ernst equation.
The choice (\ref{Z=W}) is not the unique.
We would like to emphasize here that one may construct a low-energy
string solution from every known solution of the electro-magnetic
Ernst equation in general relativity \cite{CX,Garcia}.

\subsection{Modular Transformation as T Duality}
It is naturally expected that
the modular transformation
discussed in Sect.2 can be regarded as a special
T duality. We will show that this is the case
\footnote{Although the Ernst equation is originally
an equation of the Ernst potential, which is defined in the Ehlers
picture (\ref{L_E}), we may move on to the Matzner-Misner picture
(\ref{L_MM}) through the Kramer-Neugebauer involution (\ref{KN}).
Because of this, we may well identify either modular
transformation as T duality.}.
The T-duality group has been shown to acts on $V$ as \cite{Sen2d}
\beq
V\mapsto UV,
\eeq
where
\beq
U\in
{\rm O}(8,24;\mbox{\bf Z})_L
\equiv\left\{
U\left|~U^TLU=L\right.
\right\}.
\eeq
A matrix $U$ in the form
\beq
U=\left[\begin{array}{cc}
\begin{array}{cc}U_1&U_2\\U_3&U_4\end{array}
&\\
&I_{16}
\end{array}\right] \label{U}
\eeq
belongs to $\mbox{O(8,24;{\bf Z})}_L$ if
\beqa
U_3^TU_1+U_1^TU_3=0,&&U_3^TU_2+U_1^TU_4=I_8,\n
U_4^TU_1+U_2^TU_3=I_8,&&U_4^TU_2+U_2^TU_4=0.
\eeqa
Let us now consider the subgroup
\beq
\left\{
U=\left.\left[\begin{array}{cc}\begin{array}{cc}
aI_8&bY\\cY^T&dI_8
\end{array}&\\
&I_{16}\end{array}\right]\right|~a,b,c,d\in\mbox{\bf Z},~ad-bc=1
\right\},\label{SL2Zsubgroup}
\eeq
where $Y$ is given by (\ref{XY}).
It is an elementary exercise to check that this is indeed a subgroup
of ${\rm O}(8,24;\mbox{\bf Z})_L$, and that this is isomorphic to
SL(2,{\bf Z}).

We again set $\hA_{\mu}=0$.
It is now almost clear that this SL(2,{\bf Z}) corresponds
to the integral Ehlers group of the embedded reduced Einstein
gravity. Indeed, the $V$ matrix is in the form
\beq
V=\left[\begin{array}{cc}
\begin{array}{cc}
\D^{-\frac{1}{2}}I_8&0\\\D^{-\frac{1}{2}}BY^T&\D^{\frac{1}{2}}I_8
\end{array}
&\\&I_{16}
\end{array}
\right].\label{V}
\eeq
Acting $U$ of the form (\ref{SL2Zsubgroup}) on $V$ from the left
makes $V$ deviate from the triangular gauge, so that a compensating
gauge transformation from the right is necessary.
This is the well-known nonlinear realization \cite{CCWZ}.
To be concrete, let
\beq
U=\left[
\begin{array}{cc}
\begin{array}{cc}
dI_8&cY\\bY^T&aI_8\end{array}&\\&I_{16}\end{array}
\right],~~~a,b,c,d\in\mbox{\bf Z},~ad-bc=1,
\eeq
and
\beq
h=\left[\begin{array}{cc}
\begin{array}{cc}pI_8&-qY\\qY^T&pI_8
\end{array}&\\
&I_{16}\end{array}\right],
\eeq
\beq
p=\frac{cB+d}{\left[(cB+d)^2+(c\D)^2\right]^{\frac{1}{2}}},~~
q=\frac{c\D}{\left[(cB+d)^2+(c\D)^2\right]^{\frac{1}{2}}},
\eeq
then
\beq
V\mapsto UVh
\eeq
causes the integral Ehlers SL(2,{\bf Z}) (\ref{integralEhlers}).
It is easy to check that $U_0^ThU_0$ is an element of
$\mbox{O(8)$\times$O(24)}$. This establishes the embedding of the
Ehlers SL(2,{\bf Z}) into the T duality.
Obviously, if one replaces the $V$ matrix by its image mapped by the
Kramer-Neugebauer involution, one would get an embedding of
the Matzner-Misner SL(2,{\bf Z}).
Thus either of the modular groups discussed in
the previous subsection may be identified as a subgroup of the
T duality.

\subsection{Integral
Matzner-Misner and Ehlers Symmetry as T- and S-duality}

We will now embed the Matzner-Misner SL(2,{\bf Z}) into the T-duality
in this subsection,
and will show that the Ehlers SL(2,{\bf Z}) then acts as the S-duality
on the four-dimensional sector.

We start again from the ten-dimensional action (\ref{tendimaction}).
Following Ref.\cite{MaharanaSchwarz} we introduce
the fields $\hG'_{m'n'}$, $\hB'_{m'n'}$, $\Phi'$,
$G'_{\mu'\nu'}$ and $B'_{\mu'\nu'}$ defined by
\beqa
&&\hG'_{m'n'}=G^{(10)}_{m'+3,n'+3},~
\hB'_{m'n'}=B^{(10)}_{m'+3,n'+3},~\n
&&{A'}_{\mu'}^{(m')}=\frac{1}{2}\hG'^{m'n'}G^{(10)}_{n'+3,\mu'},\n
&&{A'}_{\mu'}^{(m'+6)}=\frac{1}{2}B^{(10)}_{m'+3,\mu'}
-\hB'_{m'n'}{A'}_{\mu'}^{(n')},\\
&&G'_{\mu'\nu'}=
G^{(10)}_{\mu'\nu'}
-G^{(10)}_{m'+3,\mu'}G^{(10)}_{n'+3,\nu'}\hG'^{m'n'},\n
&&B'_{\mu'\nu'}=
B^{(10)}_{\mu'\nu'}
-4\hB'_{m'n'}{A'}_{\mu'}^{(m')}{A'}_{\nu'}^{(n')}
-2({A'}_{\mu'}^{(m')}{A'}_{\nu'}^{(m'+6)}
-{A'}_{\nu'}^{(m')}{A'}_{\mu'}^{(m'+6)}),\n
&&\Phi'=\Phi^{(10)}-\frac{1}{2}\ln\det\hG',
\eeqa
where we take $z^M$ with $M=0,\ldots,3$ as
the four-dimensional space-time coordinates $x'^{\mu'}$ $(\mu'=0,\dots,3)$,
and $z^M$ with $M=4,\ldots,9$ as the coordinates $y'^{m'}$ with
$m'=1,\ldots,6$ of the six-torus
\footnote{In this paper we use the primed indices
and quantities for the decomposition into four
and six dimensions. They appear only in this subsection.}.
The U(1)$^{16}$ gauge fields $A^{(10)}_M$ are already set to zero.
In terms of these fields the action (\ref{tendimaction}) is reduced,
by dropping the $z^{m'+3}$ dependence, to \cite{MaharanaSchwarz}
\beqa
S&=&\int d^4x'\sqrt{-G'}e^{-\Phi'}
\left[
R_{G'}+G'^{\mu'\nu'}\partial_{\mu'}\Phi'\partial_{\nu'}\Phi'\right.\n
&&\left.-\frac{1}{12}G'^{\mu'_1\mu'_2}G'^{\nu'_1\nu'_2}G'^{\rho'_1\rho'_2}
H'_{\mu'_1\nu'_1\rho'_1}H'_{\mu'_2\nu'_2\rho'_2}\right.\n
&&-G'^{\mu'_1\mu'_2}G'^{\nu'_1\nu'_2}
{F'}^{(a')}_{\mu'_1\nu'_1}(L'M'L')_{a'b'}
{F'}^{(b')}_{\mu'_2\nu'_2}\n
&&\left.
+\frac{1}{8}G^{\mu'\nu'}\mbox{Tr}(\partial_{\mu'}M'L'\partial_{\nu'}M'L')
\right],\label{action4}\\
{F'}^{(a')}_{\mu'\nu'}&=&
\partial_{\mu'}{A'}^{(a')}_{\nu'}-\partial_{\nu'}{A'}^{(a')}_{\mu'}
{}~~(a'=1,\ldots,12),\n
H'_{\mu'\nu'\rho'}&=&\left(
\partial_{\mu'}B'_{\nu'\rho'}
+2{A'}_{\mu'}^{(a')}L'_{a'b'}{F'}^{(b')}_{\nu'\rho'}\right)
+\mbox{cyclic permutations},
\eeqa
where
\beq
L'=\left[
\begin{array}{cc}
&I_6\\I_6
\end{array}
\right],~~
M'=\left[
\begin{array}{cc}
\hG'{}^{-1}&\hG'{}^{-1}\hB'\\
-\hB'\hG'{}^{-1}&\hG'-\hB'\hG'{}^{-1}\hB'
\end{array}
\right].
\eeq
To embed the Matzner-Misner SL(2,{\bf R})
we now take the ten-dimensional string-background configurations as
\beqa
&&G^{(10)}_{MN}=\left[
\begin{array}{ccc}
-\lambda'^2&&\\
&\lambda'^2&\\
&&~~~\frac{\rho}{\D}I_8~~
\end{array}
\right],
B^{(10)}_{MN}=\left[
\begin{array}{ccc}
0&&\\
&0&\\
&&
\begin{array}{cc}&B_xI_4\\
-B_xI_4&\end{array}
\end{array}
\right],\n
&&~~~~~~~~~~~~~~~~~~~~~~~~~~~~~~
e^{-\Phi^{(10)}}=\left(\frac{\rho}{\D}\right)^{-4}\rho,
\label{strbkgrd2}
\eeqa
which are the image of (\ref{strbkgrd1}) with
(\ref{Gandg}), (\ref{conformalgauge}), (\ref{hEhB})
by the Kramer-Neugebauer involution (\ref{KN}), except the conformal
factor $\lambda'$.
For our purpose we cyclically permute
the $M=3,\ldots,6$ coordinates $(3,4,5,6)\mapsto(6,3,4,5)$,
so that we have
\beqa
&&G'_{\mu'\nu'}=\left[
\begin{array}{cccc}
-\lambda'{}^2&&&\\&\lambda'{}^2&&\\
&&\frac{\rho}{\D}&\\&&&\frac{\rho}{\D}
\end{array}
\right],
B'_{\mu'\nu'}=\left[
\begin{array}{cccc}
0&&&\\&0&&\\
&&&B_x\\&&-B_x&
\end{array}
\right],
\n
&&\hG'_{m'n'}=\left[
{}~~\frac{\mbox{\normalsize $\rho$}}{\mbox{\normalsize $\D$}}I_6~~
\right],
\hB'_{m'n'}=\left[\begin{array}{cc}
&B_xI_3\\-B_xI_3&
\end{array}
\right],\n
&&H'_{\mu'\nu'\sigma'}=\partial_{\mu'}B'_{\nu'\sigma'}
+\partial_{\nu'}B'_{\sigma'\mu}
+\partial_{\sigma'}B'_{\mu'\nu'},\label{strbkgr3}\\
&&{F'}_{\mu'\nu'}^{(a')}=0~~~(a'=1,\ldots,12),\nonumber
\eeqa
and
\beq
e^{-\Phi'}=\D. \label{ePhi'=D}
\eeq
The S-duality is expressed as the invariance of the equations of
motion derived from (\ref{action4}) under the integral fractional
linear transformations of the complex potential
\beq
\Lambda\equiv\Psi+ie^{-\Phi'},\label{Lambda}
\eeq
where the scalar field $\Psi$ is defined by the relation
\beqa
H'^{\mu'\nu'\rho'}&=&
-(\sqrt{-g'})^{-1}e^{2\Phi'}\epsilon^{\mu'\nu'\rho'\sigma'}
\partial_{\sigma'}\Psi,\label{reducedH}\\
g'_{\mu'\nu'}&=&e^{-\Psi'}G'_{\mu'\nu'}.
\eeqa
Substitution of (\ref{strbkgr3}) into (\ref{reducedH}) yields
\beq
\rho^{-1}H'_{\mu 23}=-\D^{-2}\epsilon^{(2)~~\nu}_{~~~\mu}
\partial_{\nu}\Psi. \label{eqtoduality}
\eeq
Comparing (\ref{eqtoduality}) with (\ref{duality2dim}), we immediately
find that
\beq
\Psi=B
\eeq
up to a constant. Hence
$\Lambda$ defined in (\ref{Lambda}) is nothing but the Ernst potential.
Thus we see that the Ehlers SL(2,{\bf Z}) acts as the S-duality
on the four-dimensional sector
if the Matzner-Misner SL(2,{\bf Z}) is embedded into the T-duality.

\section{Colliding String Wave Solutions}
\subsection{
Ferrari-Iba$\widetilde{\mbox{\bf n}}$ez
Type Series of Solutions}

We are now in a position to utilize our results
to derive some explicit solutions of classical
string theory. Let us first consider an infinite series
of $\hB_{\mu\nu},\hA_{\mu}=0$ solutions
recast from the Ferrari-Iba$\widetilde{\rm n}$ez metrics \cite{FI}.

The Ferrari-Iba$\widetilde{\rm n}$ez metrics are given by
\beqa
ds_{\mbox{\scriptsize FI}}^2(n)&=&-\rho^{\frac{n^2-1}{2}}
(1-\xi)^{1+n}(1+\xi)^{1-n}
\left(
\frac{d\xi^2}{1-\xi^2}-\frac{d\eta^2}{1-\eta^2}
\right)\n
&&+\rho^{1-n}\frac{1+\xi}{1-\xi}dx^2
+\rho^{1+n}\frac{1-\xi}{1+\xi}dy^2,
\eeqa
where
\beqa
\xi&\equiv&u\sqrt{1-v^2}+v\sqrt{1-u^2},\n
\eta&\equiv&u\sqrt{1-v^2}-v\sqrt{1-u^2},\n
\rho^2&=&(1-\xi^2)(1-\eta^2).
\eeqa

The metric
$ds_{\mbox{\scriptsize FI}}^2(n+2)$
can be obtained from $ds_{\mbox{\scriptsize FI}}^2(n)$ by
successive applications of the Ehlers transformation (\ref{EhlersSmod})
and a change of the coordinates $x\mapsto y$, $y\mapsto -x$.
Because of the fact that
the latter interchanges the $a$- and $b$-cycles on the
``compactified'' two-torus, and the discussion in Subsect.\ref{2.2},
we see that the infinite series of metrics are generated from the
first two by
alternately applying the modular S-transformations in the
Ehlers and the Matzner-Misner pictures.
The $n$ dependence of the
conformal factor reflects the existence of the central charge
\cite{Julia} of the Kac-Moody algebra.
The embedded series of string backgrounds thus cannot be reached
each other only by T-duality transformations.

The Ernst potential of this metric is
\beq
\E_{\mbox{\scriptsize FI}}(n)=\D_{\mbox{\scriptsize FI}}(n)=
\rho^{1+n}\frac{1-\xi}{1+\xi},~~B_{\mbox{\scriptsize FI}}(n)=0.
\eeq
The conformal factor of reads
\beq
\lambda_{\mbox{\scriptsize FI}}(n)^2
=\rho^{\frac{(n+1)^2}{2}}(1-\xi)^{n+2}(1+\xi)^{-n}
\frac{8uv}{\xi^2-\eta^2}.
\eeq
Hence the conformal factor of the string solutions can be derived
by using (\ref{lambdarelation}):
\beq
\lambda^2
=uv\left(\frac{\lambda_{
\mbox{\scriptsize FI}}(n)^2}{uv}\right)^4.
\eeq
Substituting these data into (\ref{strbkgrd1}),
(\ref{Gandg}), (\ref{conformalgauge}) and (\ref{hEhB}), we get an
infinite series of the classical string solutions corresponding
to the Ferrari-Iba$\widetilde{\rm n}$ez metrics.

\subsection{$\hB_{\mu\nu}\neq 0$ Solution from a Non-collinearly
Polarized Wave}
We will next construct a less trivial, $\hB_{\mu\nu}\neq 0$ solution
from a non-collinearly polarized wave, the Nutku-Halil metric
\cite{NutkuHalil,ChandrasekharFerrari}
\beqa
ds_{\mbox{\scriptsize NH}}^2&=&
-N_0\rho^{-\frac{1}{2}}\left(
\frac{d\xi^2}{1-\xi^2}-\frac{d\eta^2}{1-\eta^2}
\right)\n
&&+\frac{\rho}{N_0}\left(
|1+p\xi+iq\eta|^2dx^2+4q\eta dx dy
+|1-p\xi-iq\eta|^2dy^2
\right)\n
&=&
-N_0\rho^{-\frac{1}{2}}\left(
\frac{d\xi^2}{1-\xi^2}-\frac{d\eta^2}{1-\eta^2}
\right)
%\n
%&&
+\frac{\rho}{N_0}\left[
\frac{N_0^2}{(1-p\xi)^2+q^2\eta^2}dx^2\right.
\n&&\left.
+\left((1-p\xi)^2+q^2\eta^2
\right)
\left(dy+\frac{2q\eta}{(1-p\xi)^2+q^2\eta^2}dx
\right)^2
\right],\label{NH}
\eeqa
where
\beq
N_0=1-p^2\xi^2-q^2\eta^2,~~~p^2+q^2=1.
\eeq
The Ernst potential $\E_{\mbox{\scriptsize NH}}$ is known to be
\cite{EGH}\footnote{Since the imaginary part of the Ernst
potential may be shifted by a constant, different expressions
appear in some literature.}
\beq
\E_{\mbox{\scriptsize NH}}=-\Xi\Pi+\frac{2(p\Pi-iq\Xi-\xi\Pi)}
{p\Xi-iq\Pi},~~~
\Xi\equiv(1-\xi^2)^\frac{1}{2},~~\Pi\equiv(1-\eta^2)^\frac{1}{2}.
\eeq
Hence
\beqa
\D_{\mbox{\scriptsize NH}}&=&\frac{\rho}{N_0}\left(
(1-p\xi)^2+q^2\eta^2
\right),\\
B_{\mbox{\scriptsize NH}}&=&
\frac{2q}{N_0}\left(
p(\xi^2-\eta^2)-\xi(1-\eta^2)
\right).
\eeqa
The conformal factor can be found as
\beq
\lambda_{\mbox{\scriptsize NH}}^2
=\rho^{\frac{1}{2}}\left((1-p\xi)^2
+q^2\eta^2\right)
\frac{8uv}{\xi^2-\eta^2}.\label{conformalNH}
\eeq
In the collinear $(q\rightarrow 0,~p\rightarrow 1)$ limit,
(\ref{NH}) is reduced to the $n=0$
Ferrari-Iba$\tilde{\rm n}$ez metric.
Using (\ref{lambdarelation}), we can similarly obtain a solution
with nonzero anti-symmetric tensor field.

\subsection{$\hB_{\mu\nu},\hA_{\mu}\neq 0$ Solution}

The final example is a $\hB_{\mu\nu},\hA_{\mu}\neq 0$ solution,
which we can construct from the Einstein-Maxwell generalization
of the Nutku-Halil metric \cite{Garcia}. In this case the Ernst and
the electro-magnetic Ernst potentials are simply given, in terms of
the Ernst potential of the Nutku-Halil metric
$\E_{\mbox{\scriptsize NH}}$, by
\beqa
\D_{\mbox{\scriptsize G}}+iB_{\mbox{\scriptsize G}}&=&
\frac{\E_{\mbox{\scriptsize NH}}}
{\psi_G\overline{\psi}_{\mbox{\scriptsize G}}},\\
E_{\mbox{\scriptsize G}}+iH_{\mbox{\scriptsize G}}&=&
\frac{(e+ib)\E_{\mbox{\scriptsize NH}}}{\psi_{\mbox{\scriptsize G}}},\\
\psi_{\mbox{\scriptsize G}}&\equiv&1+(e^2+b^2)\E_{\mbox{\scriptsize NH}},
\eeqa
where $e,b$ are arbitrary real parameters. The conformal factor is the
same as that of the Nutku-Halil metric
\beq
\lambda_{\mbox{\scriptsize G}}^2
=\lambda_{\mbox{\scriptsize NH}}^2.
\eeq
Using these data in (\ref{hEhBhA}) and (\ref{Z=W}),
we obtain a string wave solution
with non-vanishing $\hB_{\mu\nu},\hA_{\mu}$.

\section{Summary}
We have shown that the non-linear sigma models of the dimensionally
reduced Einstein and Einstein-Maxwell theories can be diagonally
embedded into the two-dimensional effective heterotic string theory.
Consequently,
the embedded string backgrounds satisfy the (electro-magnetic)
Ernst equation. In the pure Einstein theory, the Matzner-Misner
SL(2, {\bf R}) can be viewed as a change of conformal structure
of the compactified flat two-torus,
and in particular the integral ones are the modular transformations.
Hence the Ehlers SL(2, {\bf R}) and SL(2, {\bf Z}) act similarly on
another torus whose conformal structure is induced through the
Kramer-Neugebauer involution. If the Matzner-Misner SL(2,{\bf Z}) is
embedded as a special T-duality, then the Ehlers acts as the S-duality
on the four-dimensional sector. Using this embedding we constructed
some colliding string wave solutions.

\section*{Acknowledgment}
The author would like to thank H. Nicolai for valuable discussions.
This work was supported by the Alexander von Humboldt Foundation.

\section*{Note Added}
After completing the manuscript, the author became aware of
Refs.\cite{KR},
which have some overlap with the results of Sect.\ref{Embedding}
of this paper.

%\end{document}

\end{document}